\documentclass[11pt,a4paper]{article}
\usepackage{t1enc}
\usepackage[latin1]{inputenc}
\usepackage[english]{babel}
\usepackage{amssymb,latexsym,amsmath}
\usepackage{amsfonts}

\newtheorem{remark}{Remark}
\usepackage{graphicx}
\usepackage{color}

 

 \textwidth=15cm \oddsidemargin=2mm \textheight=225mm
\begin{document}

\title{\textbf{Characterization of a subclass of Tweedie distributions by a property
of generalized stability }}
\author{\textbf{L.B. Klebanov and G. Temnov}}

\author{Lev B. Klebanov  \footnotemark[1] , Grigory Temnov   \footnotemark[2] }

\footnotetext[1]{{\bf Affiliation}:\, Department of Probability and
Statistics of Charles University, Prague Sokolovska 83, Prague-8, CZ
18675, Czech
Republic\,.\,\,Email:\,klebanov@chello.cz\\
Support by the government research grant MSM 002160839 is gratefully
acknowledged}

\footnotetext[2]{Corresponding author; {\bf email:
g.temnov@ucc.ie}\,\\{\bf Affiliation}: School of Mathematical
Sciences, University
College Cork, Western Gateway Block, Cork, IE \\
\hspace*{0.45cm} }

\date{\today}\maketitle

\begin{abstract}
We introduce a class of distributions originating from an
exponential family and having a property related to the strict
stability property. Specifically, for two densities $p_\theta$ and
$p$ linked by the relation
\begin{equation*}
p_\theta(x)=e^{\theta x}c(\theta)p(x),
\end{equation*}
we assume that their characteristic functions $f_\theta$ and $f$
satisfy
\begin{equation*}
f_\theta(t)= f^{\alpha(\theta)}(\beta(\theta) t)\,\,\,\,\forall\,\,
t\in\mathbb{R},\,\,
\theta\in[a,b]\,,\,\,\,\mbox{with}\,\,\,a\,\,\,\mbox{and}\,\,\,b\,\,\,\,\mbox{s.t.}\,\,\,a\leq0\leq
b\,.
\end{equation*}
A characteristic function representation for this family is obtained
and its properties are investigated. The proposed class relates to
stable distributions and includes Inverse Gaussian distribution and
Levy distribution as special cases.

Due to its origin, the proposed distribution has a sufficient
statistic. Besides, it combines stability property at lower scales
with an exponential decay of the distribution's tail and has an
additional flexibility due to the convenient parametrization. Apart
from the basic model, certain generalizations are considered,
including the one related to geometric stable distributions.
\\[0.5cm]
\end{abstract}

\section{Introduction}

Random variables with the property of stability-under-addition
(usually called just {\it stability}) are of particular importance
in applied probability theory and statistics. The significance of
{\it stable laws} and related distributions is due to their major
role in the Central limit problem and their link with applied
stochastic models used in e.g. physics and financial mathematics.

In practical applications, stable laws appearing naturally as limit
distributions for sums of random variables (the essence of the
Central limit problem) are often used to describe the increments of
stochastic processes. Stable distributions may exhibit a reliable
fit of empirical data whose nature is related to random summation.
However, certain peculiarities of the data (e.g., lighter tails than
according to stable models) motivate the search for alternatives to
stable laws. Among different ways to adjust the tail behavior of the
stable distribution is the exponential (tilting), the smoothening of
the tail of stable density is widely applied, see e.g.
\cite{TruncL}.

In the present work, we consider exponential families with a
property that can be considered as a generalization of the usual
strict stability-under-addition property (that is, with a
real-valued function in place of the natural number of summands).
The result is, as appears, related to exponentially tilted stable
distributions.

As the class we consider originates from natural exponential
families and related to stable distributions, it is not surprising
that our class appears to be a subclass of Tweedie distributions,
that, for some combination of parameters, can also be characterized
as exponentially tilted stable distributions.

Therefore, the result of our findings is a convenient representation
of a certain subclass of Tweedie distributions -- a characterization
in terms of extended stability property. Like it is for the Tweedie
distributions, the important advantage of the proposed class is the
existence of a sufficient statistics which is due to its belonging
to natural exponential families, - and the combination of the
generalized stability with exponential tails.








\section{Definition of the class, derivation and properties of its chf }

Suppose $X$ is a real random variable with cdf $P$ and pdf $p$. A
class of distributions associated with $P$ is the {\it natural
exponential family} with the pdf
\begin{equation}\label{expFam}
p_\theta(x)=e^{\theta x}c(\theta)p(x),
\end{equation}
with the real parameter $\theta\in[a,b]$, where $a$ and $b$ are such
that $0\in[a,b]$. With $f_\theta(t)$ denoting the characteristic
function (chf) of $p_\theta(x)$ and $f(t)$ the chf of $p(x)$, the
following equality is valid according to (\ref{expFam}):
\begin{equation}\label{Eq1}
f_\theta(t)=\frac{f(t-i\theta)}{f(-i\theta)},\,\,\,\,\,
\theta\in[a,b].
\end{equation}

The problem we investigate in relation to the above chf is the
following: Do real valued functions $\alpha(\theta)$ and
$\beta(\theta)$ exist, such that
\begin{equation}\label{Eq2}
f_\theta(t)= f^{\alpha(\theta)}(\beta(\theta) t)\,\,\,\,\forall\,\,
t\in\mathbb{R},\,\,\,\,\, \theta\in[a,b]\,.
\end{equation}

It is known (see e.g. \cite{Lukacs}) that if a chf is analytic in
the neighborhood of the origin, then it is also analytic in a
horizontal strip (either this strip is the whole plane or it has one
or two horizontal boundary lines). It follows from the above that
$\forall\,\, \theta\in[a,b]$, chf $f_\theta(t)$ is analytic
($t\in\mathbb{C}$) in the horizontal strip
$\,|\,\mbox{Im}\,(t)\,|\,<\rho(\theta)$ for some $\rho(\theta)>0$.
Then with the notation $g(t):=\ln(f(t))$, one gets from (\ref{Eq1})
and (\ref{Eq2}) the following balance equation
\begin{equation}\label{Eq3}
g(t-i\theta)=g(-i\theta)+\alpha(\theta)g(\beta(\theta) t).
\end{equation}
The above functional equation can be solved using an argument based
on sequential differentiating w.r.t. to $\theta$ and to $t$, which
we postpone to Appendix. The solution w.r.t. the function $g(t)$ has
the form
\begin{equation}\label{gtSolut}
g(t)=A\left[1-(1-itc)^{\gamma}\right],
\end{equation}
where $\gamma\leq2$, $A,c\in\mathbb{R}$ and the sign of $A$ depends
on the value of $\gamma$ (as we shall see, $A<0$ when $\gamma<0$ and
when $\gamma=2$\,;\, $A>0$ when $\gamma\in(0,\,1)$ and when
$\gamma\in(1,\,2)$). The expression for the chf $f(t)$ immediately
follows
\begin{equation*}\label{ftExp}
f(t)=\exp\left\{A\left[1-(1-itc)^{\gamma}\right]\right\}.
\end{equation*}

Recalling (\ref{Eq1}), we can now write down the chf $f_\theta$
\begin{eqnarray}\label{ftTheta2}
f_\theta(t)&=&\exp\left\{A\left[(1-c\theta)^{\gamma}-(1-itc-c\theta)^{\gamma}\right]\right\}
\\
&=&\exp\left\{A\cdot(1-c\theta)^{\gamma}\left[1- \left(1-
\frac{ itc }{1-c\theta } \right)^{\gamma}\right]\right\}\nonumber \\
&=& \exp\left\{A\cdot B^{\gamma}\left[1- \left(1- \frac{ itc }{B }
\right)^{\gamma}\right]\right\},
\end{eqnarray}
where $B:= 1-c\theta$.

Note that the functions $\alpha(\theta)$ and $\beta(\theta)$ are
then explicitly expressed
\[ \alpha(\theta)=(1-c\theta)^\gamma\,,\,\,\,\,  \beta(\theta)=1/(1-c\theta)\,. \]

\subsection{ Ranges of the shape exponent $\gamma$; role of parameters $A$ and $c$ }\label{cases}

The properties of introduced distributions differ with respect to
various combinations of their parameters $\gamma,A,c$.

As the role of the shape exponent $\gamma$ is crucial to relations
between other parameters, in the present paragraph we analyze
different cases that arise with respect to certain ranges of
$\gamma$. That classification w.r.t. $\gamma$ also makes it easier
to clarify the relation of the introduced class to stable laws and
other important distributions.


Returning to the representation for the chf $f$
\begin{equation}\label{basicf}
f_\gamma(t)=\exp\left\{A(\left[1-(1-itc)^{\gamma}\right]\right\}\,,
\end{equation}
note that the properties of the corresponding distribution differ
depending on the range of $\gamma$ (to mark the importance of this
dependency, we shall use the notation $f_\gamma$, instead of just
$f$, and will point out the range of $\gamma$, such as
$f_{\gamma<0}$ or $f_{\gamma\in(1,2)}$). Listing different cases
w.r.t. $\gamma$ below, we shall see that for each of the ranges of
$\gamma$, $f_\gamma$ is a proper chf only for a particular sign of
$A$. In all cases, there exists a corresponding analytic function
$f_\gamma(t),\,\,t\in\mathbb{C}$, that agrees with the
characteristic function $f_\gamma(t),\,\,t\in\mathbb{R}$. The domain
of regularity for corresponding analytic function varies depending
on the ranges of $\gamma$ considered below.

\paragraph{ Case {\bf a}\,:\, $\gamma<0$\,} Denoting $\overline{\gamma}=-\gamma$, for this case we have
\begin{equation}\label{case-a}
f_{\gamma<0}(t)=\exp\left\{A\left[1-(1-itc)^{
-\overline{\gamma}}\right]\right\}.
\end{equation}
As $\widehat{\gamma}(t):=(1-itc)^{ -\overline{\gamma}}$ is the chf
of a gamma distribution, we see that $f_{\gamma<0}$ corresponds to
the chf of the compound Poisson r.v. with gamma-distributed
summands, provided that $\overline{A}:=-A>0$:
\begin{equation}\label{case-a-1}
f_{\gamma<0}(t)=\exp\left\{\overline{A}\left[\widehat{\gamma}(t)-1\right]\right\}.
\end{equation}
In other words, $f_{\gamma<0}(t)$ is a chf of the random sum
$S=\sum\limits_{i=1}^{N(t)}X_i$, where $N\sim Poisson$ and $X_1\sim
gamma(\overline{\gamma},c)$\,, i.e. $\overline{\gamma}$ and $c$ are
the shape and the scale parameter correspondingly.

Clearly, $f_{\gamma<0}(t), \,\,t\in\mathbb{C}$,\,\,is analytic when
$\mbox{Im}(t)\in\left(-1/c,\infty\right)$, with the entire line
$t=-i/c$ being the region of singularity for $f(t)$.

\paragraph{ Case {\bf b}\,:\, $\gamma\in(0,1)$\,} This case, after a complex shift of the variable,
corresponds to a certain subclass of stable distributions.
Specifically, recall that if $\widetilde{X} = X+x_0$ is a {\it
shifted one-sided Levy-stable} variable then its chf can be
represented as
\begin{equation}\label{Sh1Sd}
\widetilde{\phi}(z)=\exp\left\{izx_0 -
(-iz)^\alpha\widetilde{a}\right\},\,\,\,\,\,\widetilde{a}>0.
\end{equation}
Clearly, (\ref{Sh1Sd}) with $x_0=0$ brings us close to our case
(\ref{basicf}), differing by just a proper change of the variable.
Specifically, with $cz:=ct+i$ we have
\begin{equation}\label{case-b}
f_{\gamma\in(0,1)}(t)=e^{A}\exp\left\{-
(1-itc)^{\gamma}A\right\}\,\,\,\Longleftrightarrow\,\,\,f_{\gamma\in(0,1)}(z-i/c)=e^{A}\exp\left\{-
(-izc)^{\gamma}A\right\}\,,
\end{equation}
which is close to (\ref{Sh1Sd}) up to the normalization factor $e^A$
(provided that $A>0$).

Furthermore, note that the re-parametrization $cz=ct+i$ relates to
the transformation of the density, similar to the one used in the
introduction of $f_\theta$, according to (\ref{expFam}) and
(\ref{Eq1}). When $t=z-i/c$ then in order for
$f_{\gamma\in(0,1)}(t)$ to correspond to a proper chf we need a
normalization factor $f_{\gamma\in(0,1)}(-i/c)$.
From (\ref{case-b}), we see that
$\frac{f_{\gamma\in(0,1)}(z-i/c)}{f_{\gamma\in(0,1)}(-i/c)}=:f_{\gamma}^{stab}(z)$
corresponds to one-sided stable chf and that
$f_{\gamma\in(0,1)}(-i/c)=e^A$.

Hence, one-sided stable density corresponding to chf
$f_{\gamma}^{stab}$ is the result of an exponential transformation
(take the exponent parameter $1/c$ in (\ref{expFam}) instead of
$\theta$) of the density corresponding to chf $f_{\gamma\in(0,1)}$.
That is, the r.v. with chf $f_{\gamma\in(0,1)}$ in turn relates to
the exponential transformation of the one-sided stable density with
the exponent parameter being $-1/c$. Note that both
$f_{\gamma}^{stab}$ and $f_{\gamma\in(0,1)}$ are proper
characteristic functions only when $A>0$.

Like in Case {\bf a}, the function
$f_{\gamma\in(0,1)}(t)\,,\,t\in\mathbb{C}\,,$ is analytic in
$\mbox{Im}(t)\in(-1/c,\infty)$, but contrary to the previous case,
there exists a limit for $f_{\gamma\in(0,1)}(t)$ w.r.t $t$
approaching the line $t=-i/c$ (with the only point of non-regularity
being $(0,-i/c)$). This limit corresponds to $f_{\gamma}^{stab}$
introduced above which is the chf of one-sided stable law.

Note that the above is in correspondence with one of the classical
results on characteristic functions that says that a necessary
condition for a function analytic in some neighborhood of the origin
to be a characteristic function is that in either half-plane the
singularity nearest to the real axis is located on the imaginary
axis (see e.g. \cite{Lukacs} for details).

\begin{remark}
Check that the well-known {\it Inverse Gaussian distribution} as a
special case with $\gamma=1/2$.
\end{remark}


\paragraph{ Case {\bf c}\,:\, $\gamma=1$\,} This degenerate case corresponds to the constant
r.v. whose cdf is the Heaviside step function, since the
distribution corresponding to the chf $e^{iAct}$ has a single unit
jump at point $Ac$.

\paragraph{ Case {\bf d}\,:\, $\gamma\in(1,2)$\,}

Consider a re-parametrization $\widetilde{\gamma}=\gamma/2$ allowing
to write down the chf $f(t)$ as
\begin{equation}\label{1stmix}
\widetilde{f}_{\gamma\in(1,2)}(t) = \exp\left\{A[1-
((1-itc)^2)^{\widetilde{\gamma}}]\right\}=\exp\left\{A[1-
(1-c\widetilde{t})^{\widetilde{\gamma}}]\right\}=f_{\widetilde{\gamma}\in(0,1)}(-i\widetilde{t}),
\end{equation}
where $\widetilde{t}:=2it+ct^2$\,.

While $f_{\widetilde{\gamma}\in(0,1)}(t)$ is a chf of the type
considered above (Case {\bf b}), $f_{\gamma\in(1,2)}(t)$ is also a
chf. Indeed, while $\phi_{(\mu,\sigma)}(t)=\exp\left\{\mu
it-\sigma^2t^2/2\right\}$ is the chf of the Gaussian r.v. (with
expectation $\mu$ and variance $\sigma^2$), the function
\begin{equation}\label{LaplMix}
f_{\widetilde{\gamma}\in(0,1)}(-i\widetilde{t})=\int
e^{\widetilde{t} x}p_{\widetilde{\gamma}}(x)dx = \int
\left[\phi_{(2,2\overline{c})}(t)\right]^x
p_{\widetilde{\gamma}}(x)dx
\end{equation}
which is a chf of some probabilistic distribution, as a continuous
mixture of the Gaussian chf
$\phi_{(2,2\overline{c})}(t)=\exp\left\{2
it-\overline{c}t^2\right\}$ (provided that $\overline{c}:=-c>0$)
with the pdf $p_{\widetilde{\gamma}}$ whose chf
$f_{\widetilde{\gamma}\in(0,1)}$ is the one from the above Case {\bf
b}.

Since we already know from Case {\bf b} that it should be that $A>0$
for $f_{\gamma\in(0,1)}$ to be a chf, in the present case we also
have $A>0$ as a necessary condition for $f_{\gamma\in(1,2)}$ given
by (\ref{LaplMix}) to be a chf. Like in Case {\bf b}, $f(t)$,
$t\in\mathbb{C}$, is analytic only in the strip restricted by the
horizontal line $t=-i/c$, i.e. when $\mbox{Im}(t)\in(-1/c,\infty)$,
but additionally we have the condition that $c<0$.

\paragraph{ Case {\bf e\,}:\, $\gamma=2$\,} Check that $f_{\gamma=2}$ is the chf of the
normal r.v.
\begin{equation}\label{chfnorm}
f_{\gamma=2\,}(t)=\exp\left\{\overline{A}\left(-2cit -
c^2t^2\right)\right\}=\phi_{(-2\overline{A}c,2\overline{A}c^2)}(t)\,,
\end{equation}
where $\overline{A}=-A$. As the variance is
$\sigma^2=2\overline{A}c^2$, it should be that $A<0$.

\vspace*{0.4cm}

\paragraph{The sign of $c$}

Discussing the cases with respect to the parameter $\gamma$, we
noted that the sign of the parameter $A$ is crucial in each of the
cases. Commenting on the sign of the other parameter, $c$, we noted
that its sign is only crucial in Case {\bf d}. In other cases, $c$
can take any sign when $f_\gamma$ is a proper chf due to the
property of characteristic functions providing that $f(-t)$ is a chf
as soon as $f(t)$ is a chf. However, we should keep in mind that the
function $f_\gamma(t)$ ($t\in\mathbb{C}$), in all relevant cases, is
analytic only in a strip depending on $c$.

\paragraph{Limit behavior of $f$ regarding mutual limit w.r.t. $c$ and $A$}

Note that the one-sided stable distributions appearing in Case {\bf
b} and obtained via the change of the complex variable $cz:=ct+i$
could also be viewed as a limiting case w.r.t $c\rightarrow\infty$
and $A\rightarrow0$. Additionally, $c$ and $A$ should be linked so
that $A \sim
\widetilde{c}/c^\gamma$\,\,\,\,($\widetilde{c}\in\mathbb{R}$), hence
the limit w.r.t $A\rightarrow0$ and $c\rightarrow\infty$ (keeping
$\widetilde{c}$ constant) leads to
\[ f_{\gamma\in(0,1)}(t)=e^{A}\exp\left\{-
\left(\frac{1}{c}-it\right)^{\gamma}\left(c^{\gamma}A\right)\right\}
\xrightarrow[\underset{A\rightarrow0}{c\rightarrow\infty}]{}{\exp\left\{-
(-it)^{\gamma}\widetilde{c}\right\}}=:f^{(lim)}_{\gamma\in(0,1)}(t),
\] which corresponds to one-sided stable distributions appearing in
Case {\bf b} above. Recalling the notation of Case {\bf b}, we see
that $f^{(lim)}(t)=f_{\gamma}^{stab}(t)=\exp\left\{-
(-itc)^{\gamma}A\right\}$.

\subsection{ The role of the natural exponent $\theta$ }\label{Lim}

Let us turn again to the representation for $f_\theta$. Rewrite
(\ref{ftTheta2}) noting that an additional re-parametrization with
$\widetilde{A}:=A\cdot B^{\gamma}$ and
$c_\theta=\frac{c}{1-c\theta}$ leads to the form (\ref{basicf})
\begin{eqnarray}\label{ftTheta3}
f_{\gamma,\theta}(t)=\exp\left\{\widetilde{A}\cdot\left[1-(1-itc_\theta)^{\gamma}\right]\right\},
\end{eqnarray}
where we used the notation $f_{\gamma,\theta}$ instead of just
$f_\theta$, indicating the significance of both of the parameters.

Clearly, {\it Cases} considered above in relation to chf $f_\gamma$
can be revisited for $f_\theta$, but apart from the ranges of the
shape exponent $\gamma$, the role the natural exponent $\theta$ is
important.


Obviously, the borders of the analyticity strips depending on $c$,
as well as the conditions on $A$, will change in correspondence with
$\theta$, i.e. one can view introducing $f_{\gamma,\theta}$ in place
of $f_\gamma$ as passing from $\theta = 0$ to non-zero $\theta$, so
that the "new" $c$ and $A$ will depend on $\theta$.


In particular, replacements through Case {\bf a} -- Case {\bf e}\,
are the following:
\begin{itemize}
\item
the strips of analyticity: $(-i/c,\infty)\Longrightarrow
\left(-i(1/c-\theta),\infty\right)$ in Cases {\bf a}, {\bf b} and
{\bf d};
\item
the signs of coefficients: recall that Case {\bf a} and Case {\bf e}
imply $A<0$, while in both Case {\bf b} and Case {\bf d} we have
$A>0$; for $f_{\gamma,\theta}$ the same holds with $A\Longrightarrow
A\cdot B^\gamma = A\cdot(1-c\theta)^\gamma$.
\end{itemize}

Revisit, for instance, Case {\bf d} where we have a representation
analogous to the one expressed by (\ref{1stmix}) and
(\ref{LaplMix}):
\begin{equation}\label{LaplMixN}
\widetilde{f}_{\gamma\in(1,2),{\theta}}(t)
=f_{\widetilde{\gamma}\in(0,1),{\theta}\,\,}(2t-ict^2) = \int
\left[\phi_{(2,2\overline{c}_{\theta})}(t)\right]^x
p_{\widetilde{\gamma},\theta}(x)dx\,,
\end{equation}
where pdf $p_{\widetilde{\gamma},\theta}$ corresponds to the chf
$f_{\widetilde{\gamma}\in(0,1),{\theta}}=\exp\left\{\widetilde{A}\left[1-
(1-itc_{\theta})^{\widetilde{\gamma}}\right]\right\}$ and
$\phi_{(2,2\overline{c}_{\theta})}$ is a chf of the normal
r.v.,\,\,i.e.\, $\phi_{(2,2\overline{c}_{\theta})}(t)=\exp\left\{2
it-\overline{c}_{\theta}t^2\right\}$ (with
$\overline{c}_{\theta}:=-c_{\theta}$). So that in order for
$\widetilde{f}_{\gamma\in(1,2),{\theta}}$ to be a chf, it should be
that $c_{\theta}=c/(1-c\theta)<0$ and $\widetilde{A}=A\cdot
B^{\gamma}>0$.

All other cases w.r.t $\gamma$ can be revisited for
$f_{\gamma,\theta}$ analogously. Additionally, certain values and
ranges of $\theta$ correspond to particular special cases which we
consider below.

\paragraph{Range of $\theta$, singularities and corresponding limits.}

While $c$ can be positive or negative, the point when $\theta=1/c$
is the case of singularity and/or the limiting case.

Specifically, Case {\bf a} with $\theta=1/c$ is the case of
singularity of the chf $f_{\gamma,\theta}$. In Case {\bf b}, the
limit $1/c\leftarrow\theta$ exists with limiting chf's
$f^{(lim)}_{\gamma}(t)=\exp\left\{- (-izc)^{\gamma}A\right\}$.


In a way, approaching the point $1/c$ with respect to \, $\theta$ \,
means approaching corresponding stable (or mixed normal-stable)
distributions, while increasing the exponent parameter $\theta$
corresponds to lightening the tails of distributions $P_\theta$ to
exponential ones.



\paragraph{Limit case $\theta\longrightarrow\infty$\,.}

The question in interest is: what distributions appear in the limit
case $\theta\longrightarrow\infty$? In \cite{LimExp}, limit laws for
the whole class of natural exponential families w.r.t. the growing
exponent parameter were investigated. It was shown that for the r.v.
$X_\theta$ with density of the form (\ref{expFam}), constants
$a_\theta>0$ and $b_\theta$ can chosen so that a limit
$(X_\theta-b_\theta)/a_\theta \longrightarrow Y,\,\,
\theta\rightarrow\infty$ exists and that only possible limit
distributions $Y$ are the normal and gamma distributions.
Specifically, the following results were proved. {\it
\begin{itemize}
\item If $P_\theta$, $\theta\in\Theta$, is the cdf of the exponential
family $X_\theta$ whose pdf is given by (\ref{expFam}) and if there
exist constants $a_\theta >0$ and $b_\theta \in\mathbb{R}$ such that
$P_\theta\left(\frac{x-b_\theta}{a_\theta}\right)\xrightarrow
[\theta\rightarrow\infty]{}G$ weakly to some non-degenerate cdf $G$
then $G$ belongs to the so called {\it extended gamma family}, which
includes normal and gamma densities.
\item In the case of convergence to a non-degenerate limit $G$,
$a_\theta$ and $b_\theta$ should be taken such that $P_\theta$ is
centered and scaled by expectation and standard deviation.
\end{itemize}
}

Criterion of the convergence of an exponential family (\ref{expFam})
to gamma/normal distribution can be expressed in terms of the {\it
moment generating function} (mgf) associated with $p(x)$, i.e.
$M(\lambda)=\int e^{\lambda x} p(x)dx$. Specifically, as proved in
\cite{LimExp}, the following holds.

{\it If the function
$s(\lambda)=1/\sqrt{m^{\prime\prime}}(\lambda)$, where
$m(\lambda):=\ln M(\lambda)$, is self-neglecting, then the
exponential family $X_\lambda$, $\lambda\in\Theta$, is
asymptotically normal. Otherwise, the only possible limit
distribution is gamma distribution.}

It is easy to check when the normal distribution appears as a limit
in our case. Recalling (\ref{gtSolut}), for the log--mgf (or
cumulant generating function) of the introduced family we have
\begin{equation*}
m(\lambda)=A\left[1-(1-\lambda c)^{\gamma}\right],
\end{equation*}
so that
\begin{equation*}
m^{\prime\prime}(\lambda)=Ac^2\gamma(\gamma-1)\left[(1-\lambda
c)^{\gamma-2}\right].
\end{equation*}
Clearly, the above condition on $s=1/\sqrt{m^{\prime\prime}}$ is
only satisfied when $\gamma\geq2$. That means that in all the cases
w.r.t. to $\gamma$ that are relevant for our study, the only
possible  limit distribution $G$ is gamma distribution, i.e.
according to the usual terminology, our distribution lies in the
{\it domain of attraction} ($DA$) of gamma distribution.

\section{Relation to the Tweedie class and special cases}

Like briefly noted above, the proposed distributional family
includes some well-known models, which can be considered as special
cases of exponential family with stability property. At the same
time, it itself is subclass of another class -- the Tweedie
distributions \cite{Tweed}, which, in its turn, is a part of even a
wider class of {\it exponential dispersion models}.

\subsection{Relation to the Tweedie class}

For general definitions and characterizations of {\it exponential
dispersion models} we refer to \cite{Jorgens}.
The pdf of an exponential dispersion model (EDM) is represented in a
way similar to that of our class
\[
f_{ED}(y)=p_{\beta}(y)\exp\left\{\frac{1}{\beta}(\theta
y-\kappa(\theta))\right\}\,,
\]
where $\beta>0$. A convenient representation of the Tweedie class is
through the moment generating function $M(t)=\int\exp(ty)f(y)dy$.
The cumulant generating function of the EDM
 is
\[ m_{ED}(t)=\log M_{ED}(t)=[\kappa(\theta+t\phi)-\kappa(\theta)]/\beta\,. \]
As the derivatives of the function $\kappa(\cdot)$ w.r.t $\theta$
provide the values of the cumulants of the distribution,
$\kappa(\cdot)$ is called the cumulant function. So that we have
$\mu=\kappa^{\prime}(\theta)$ for the mean and
$\beta\kappa^{\prime\prime}(\theta)$ for the variance. Since the
mapping from $\theta$ to $\mu$ is invertible, there exists a
function $V(\cdot)$, called the variance function, such that
$\kappa^{\prime\prime}(\theta)=V(\mu)$.

The Tweedie class is then defined as EDM with the variance function
$V(\cdot)$ of a specific form, namely $V(\mu)=\mu^p$ for
$p\in(-\infty,0]\cup[1,\infty)$ and $V(\mu)=\exp(\upsilon\mu)$ for
$p=\infty$ (here, $\upsilon\neq0$). The correspondent cumulant
function $\kappa$ for the Tweedie class can be represented as
\begin{equation}\label{S1}
\left\{
\begin{array}{rcl}
\frac{\mu^{2-p}-1}{2-p}\,\,,\,\,\mbox{when}\,\,\,\, p&\neq&2\,\\
\log\mu \,\,,\,\,\mbox{when}\,\,\,\,p&=&2\,.
\end{array}
\right.
\end{equation}
Defining a new parameter $\alpha$ via
\[ (p-1)(1-\alpha)=1\,, \]
a representation for the moment generating function of the Tweedie
class is obtained:
\begin{equation}\label{S2}
M_{Tw}(s)=\left\{
\begin{array}{rcl}
\exp\left\{\frac{(1-p)^\alpha\theta^\alpha}{\beta(2-p)}
\left[(1+s/\theta)^\alpha-1\right]\right\} \,\,,\,\,\mbox{when}\,\,\,\, p&\neq&1,2\,,\\
(1+s\theta)^{-1/\beta} \,\,,\,\,\mbox{when}\,\,\,\, p&=&2\,,\\
\exp\left\{\frac{1}{\beta} e^{\theta}(e^s-1)\right\}
\,\,,\,\,\mbox{when}\,\,\,\, p&=&1\,.
\end{array}
\right.
\end{equation}

From above, comparing with the chf (\ref{basicf}) of our class, it
is clear that our class is actually a subclass of Tweedie
distributions with some parameter $p\neq1,2$ and a certain
combination of other parameters.

Correspondingly, the basic properties that we discussed above do
hold for the Tweedie class as well. For instance, it is known that
the Tweedie distributions can be obtained through exponential
damping of stable distributions.

The distributions of the Tweedie class are known to have a {\it
scale invariance property}, that can be expressed through the pdf
$f_{Tw}$ as
\[ f_{Tw}(x;\mu,\phi)=cf_{Tw}(cx;c\mu,c^{2-p}\beta)\,,\,\,\,\,\mbox{for}\,\,c>0\,, \]
which also resembles stability property (in terms of scale
invariance), however is not directly related to the generalized
stability property introduced by considering the relation
(\ref{Eq2}).

\vspace*{0.2cm}

Note also that our subclass does not explicitly contain the
parameter $\beta$ in the above ED notation (parameter related to the
variance of the distribution).

Besides, note that gamma distribution belonging to the Tweedie
class, in our case appears as a limit case
$\theta\longrightarrow\infty$, like mentioned in the previous
section.





\subsection{Inverse Gaussian and Levy distributions}

Two important special cases are visible immediately from the form of
the characteristic functions $f_{\gamma}$ and $f_{\gamma,\theta}$.

Recall the characteristic function representation of the well-known
{\it inverse Gaussian distribution}
\begin{equation}\label{IGD}
g_{IG}(t)=e^{\frac{\lambda}{\mu}\cdot\left[1-\left(1-\frac{2\mu^2it}{\lambda}\right)
^{1/2}\right]}
\end{equation}
and note (as already briefly remarked in Case {\bf b} \,of Paragraph
\ref{cases}) that it can be viewed as a special case of the
introduced exponential family with stability
--- just plug the parameters' values
$(A=\frac{\lambda}{\mu},c=\frac{2\mu^2}{\lambda},\gamma=\frac{1}{2})$
into the general representation (\ref{basicf}).

Recall an important property of the inverse Gaussian distribution:
\begin{itemize}
\item If $X_i$ has an $IG(\mu w_i,\lambda w^2_i)$ distribution for
$i=1,2,\dots,n$ and $X_i$ are independent, then
$S=\sum_{i=1}^{n}X_i\sim IG(\mu \overline{w},\lambda
\overline{w}^2)$, where $\overline{w}=\sum\limits_{i=1}^n w_i$. In
other words, $X\sim IG(\mu,\lambda)\quad\Rightarrow tX\sim
IG(t\mu,t\lambda)$
\end{itemize}
Note that this property is related to the scale invariance property
and recall that stable distributions are also included in certain
subclasses of introduced exponential family (see Case {\bf b} and
Case {\bf d}), so that the scale invariance appears to be naturally
embedded in our model.

Another well known distribution appears as a limiting case of the
Inverse Gaussian distribution: the chf $f_{Levy}$ of the {\it Levy
distribution} is clearly a limit of $g_{IG}$ w.r.t.
$\mu\longrightarrow\infty$ (and respectively
$\lambda\longrightarrow0$) which leads from (\ref{IGD}) to
\begin{equation}\label{levvy}
f_{Levy}(t)=e^{-\sqrt{-2i\widetilde{c}t}}\,.
\end{equation}
Inverse Gaussian and Levy distributions are the important cases as
they have simple explicit forms for their pdf's.


\subsection{ Note on Tempered Levy processes}

The combination of the properties of our subclass relates in its
purpose to other modified models such as the so-called Tempered Levy
process.


The idea behind Tempered Levy processes (also known as Truncated
Levy flights \cite{TruncL}) is to use a distribution which coincides
with the stable one around zero (i.e. for small fluctuations) and
has heavy tails yet decreasing to zero fast enough to assure a
finite variance.

The latter is achieved by truncating stable distributions, and one
of the ways the truncation can be done is the so called smooth
exponential truncation. The result  corresponds to an infinitely
divisible distribution enabling an analytical chf representation:
\[
\phi_{\alpha;\mu}(t)=\exp\left\{\frac{-a_\alpha}{\cos(\pi\alpha/2)}
\left[(\mu^2+t^2)^{\alpha/2} \cos\left(\alpha
\arctan\left(\frac{|t|}{\mu}-\mu^\alpha\right)\right)\right]\right\},
\]
where $\mu$ is the cut-off parameter. The asymptotic behavior of
$\phi$ is given by
\[ \phi_{\alpha;\,\mu}(t)\underset{t\rightarrow\infty}
{\sim}\phi_\alpha(t) \] where $\phi_\alpha(t)$ is the chf of the
(non-truncated) stable distribution, i.e. for small values of $x$,
the pdf $p_{\alpha;\mu}(x)$ corresponding to the chf
$\phi_{\alpha;\mu}$ behaves like a stable law of index $\alpha$.

This concept is then interpreted in terms of Levy processes and
associated Levy measures. Specifically, if $\nu(x)$ is a measure of
a Levy process (generally, defined as the expected number, per unit
time, of jumps whose size is less than $x$), then the measure
\[ \widetilde{\nu}(dx)=e^{-\theta x}\nu(dx), \]
corresponds to another Levy process, whose large sizes are
"tempered" with exponential damping.

The ideas are therefore similar to the one that lead to the
introduction of the Tweedie class, as well as to the
characterization of our subclass, yet with certain methodological
difference: The idea of exponential tempering utilizes stable
distribution as the initial distribution whose tail is then
exponentially smoothened, while in our approach the stability
property appears naturally from the properties of the introduced
class. That allows to view the exponential family with stability
property as a natural extension of both classes: stable
distributions and exponential families which is promising regarding
their practical applications, including the use of sufficient
statistic naturally inherent in exponential families.

\section{Modifications related to geometric stable laws }\label{GSect}

\paragraph{Preliminaries and definitions}

First, recall the origin of {\it geometric infinitely divisible}
(GID) distributions.

The chf of an infinitely divisible (ID) r.v. $X$ implies the
representation $\phi(t) = (\phi_n(t))^n$, where the chf $\phi_n(t)$
corresponds to some r.v. $X^{(n)}$, which in turn implies that the
sum $X_1^{(n)}+\dots+X_n^{(n)}$ of iid copies of $X^{(n)}$ has the
same distribution as $X$. The so-called {\it transfer theorems} (see
e.g. \cite{GnedKor}) state that the random sums
\begin{equation}\label{sums}
X_1^{(n)}+\dots+X_{\nu_n}^{(n)},
\end{equation}
where $(\nu_n)$ is a sequence of integer-valued r.v.'s such that
$\nu_n\stackrel{p}{\longrightarrow}\infty$ (in probability) while
$\nu_n/n\stackrel{d}{\longrightarrow}\nu$ (in distribution),
converge (in distribution) to a r.v. $Y$ whose chf has the form
\begin{equation}\label{LS}
\omega(t) = \mathcal{L}(- \ln \phi(t)),
\end{equation}
where $\mathcal{L}$ is the Laplace transform of the r.v. $\nu$.

If $\nu_n\stackrel{d}{=}\nu_p$ has a geometric distribution with
mean $1/p$ (where $p$ is close to zero, and $\nu_p$ is independent
of $X_i's$) which converges in distribution to the standard
exponential distribution with Laplace transform
$\mathcal{L}(u)=(1+u)^{-1}$ then (\ref{LS}) turns into
\begin{equation}\label{GS}
\omega(t) = (1 - \ln \phi(t))^{-1},
\end{equation}
and random variables whose chf has the form (\ref{GS}) were
introduced in \cite{KlebanovMM} as {\it geometric infinitely
divisible} (GID) random variables. Since sums such as (\ref{sums})
frequently appear in many applied problems, GID have a variety of
applications. The {\it geometric stable distributions} (GS), also
originated from \cite{KlebanovMM} and developed in later works e.g.
\cite{GnedKor} and \cite{HeavyTailed}, appear as a natural subclass
of GID and are widely used for modeling of stochastic processes. The
chf of GS laws has the form (\ref{GS}) with $\phi(t)$ being the chf
of a stable r.v.


It is natural to extend the exponential family with stability
property introduced in previous sections w.r.t geometric stability.
Specifically, for the chf $f_{\gamma,\theta}$ of the form
(\ref{ftTheta3}), the corresponding geometric extension will have
the chf
\begin{equation}\label{EqN}
\omega_{\gamma,\theta}(t)= \frac{1}{1-\ln
f_{\gamma,\theta}(t)},\,\,\,\forall \,\,\,t\in\mathbb{R},\,\,\,\,\,
\theta\in[a,b]\,.
\end{equation}
Since
$f_{\gamma,\theta}(t)=\exp\left\{\widetilde{A}\cdot\left[1-(1-itc_\theta)^{\gamma}\right]\right\}$,
 note that when $\widetilde{A}=A\cdot B^{\gamma}<0$ (which
corresponds to certain ranges of $\gamma$) then (\ref{EqN}) can
viewed as a sum of geometric progression with the initial value
$(1-\widetilde{A})^{-1}$ and common ratio
$\frac{\widetilde{A}}{1-\widetilde{A}}\left(1-
itc_\theta\right)^{\gamma}$.

Revisiting the cases considered in Paragraph \ref{cases} again, this
time in relation to $\omega_{\gamma,\theta}$, the argument above
leads to a convenient interpretation of Case {\bf a}, which we
sketch below, along with other cases w.r.t the ranges of $\gamma$.

\begin{itemize}
\item {\bf Case  a} $\gamma<0$.\,\,In this case, $f_{\gamma<0,\theta}$ is a chf of a
probability distribution only when $\widetilde{A}<0$, which
coincides with the condition under which the representation
(\ref{EqN}) can be interpreted as a geometric progression. Thus the
cdf \,$\Omega_{\gamma<0,\theta}$ corresponding to the chf
$\omega_{\gamma<0,\theta}$ has the representation as a series of
convolutions of the gamma distribution with itself:
\begin{equation}\label{InfSum}
\Omega_{\gamma<0,\theta}(x)=
\frac{1}{1-\widetilde{A}}\sum\limits_{n=0}^{\infty}
\left(\frac{\widetilde{A}}{1-\widetilde{A}}\right)^n
\overline{F}_{\overline{\gamma},\theta}^{*n}(x),
\end{equation}
where
$\overline{F}_{\overline{\gamma},\theta}(x)=1-F_{\gamma,\theta}(x)$,
and $F_{\overline{\gamma},\theta}$ is the cdf of gamma distribution
with shape parameter $\overline{\gamma}=-\gamma$ and scale parameter
$c_\theta$.




\item {\bf Case b} $\gamma\in(0,1)$. In this case $f_{\gamma\in(0,1),\theta}$ is a chf only when
$\widetilde{A}>0$, so the representation of the form (\ref{InfSum})
is not valid. However, as the representation (\ref{EqN}) implies
that any $\omega_{\gamma,\theta}$ is a chf as soon as
$f_{\gamma,\theta}$ is a chf, it follows that
$\omega_{\gamma\in(0,1),\theta}$ is the chf (provided that
$\widetilde{A}>0$), and the corresponding r.v. is a geometric
analogue of exponentially transformed one-sided stable r.v.,
corresponding to the chf $f_{\gamma\in(0,1),\theta}$.


\item {\bf Case d}\,\, $\gamma\in(1,2)$. Recall the representation
(\ref{LaplMix2}) meaning that $f_{\gamma\in(1,2),\theta}$ is a chf
linked via the transformation $\int
\left[\phi_{(2,2\overline{c}_{\theta})}(t)\right]^x
p_{\widetilde{\gamma},\theta}(x)dx$ with the pdf
$p_{\widetilde{\gamma},\theta}$ of Case {\bf b}
($\widetilde{\gamma}\in(0,1)$) and with the chf
$\phi_{(2,2\overline{c}_{\theta})}$ of the normal distribution (with
mean and variance $(2,2\overline{c}_{\theta})$). Then
$\omega_{\gamma\in(1,2),\theta}$ given by (\ref{EqN}) is also a chf,
the geometric analogue of $f_{\gamma\in(1,2),\theta}$, provided that
$\widetilde{A}>0$ and additionally $c_\theta<0$.

\item {\bf Case e}\,\, $\gamma=2$. As chf $f_{\gamma=2,\theta}$
corresponds to the Normal r.v., $\omega_{\gamma=2,\theta}$ is the
chf of the geometric analogue of the Normal r.v.

Check that this case can also be interpreted in the way similar to
the Case {\bf b}, so that $\omega_{\gamma=2,\theta}$ can be viewed
as an exponentially transformed Laplace r.v.


\item {\bf Case f}\,\, $\gamma>2$. This is a new case in our
framework, since $f_{\gamma>2,\theta}\,(t)$ \,\, does not correspond
to a proper probability distribution (in the representation
$\int\left[e^{P(t)}\right]^x p_\theta(x)dx$ that we used in Case
{\bf b} and Case {\bf d}, the polynomial $P(t)$ can not have a
degree greater than $2$, due to Marcinkiewicz's theorem
\cite{Marcin}). \par However, $\omega_{\gamma>2,\theta}(t)=(1-\ln
f_{\gamma>2,\theta}(t))^{-1}$ {\it is} a proper chf. That could be
seen if we assume $\gamma\in(2,4)$, denote
$\widetilde{\gamma}:=\gamma/2$ and $a:=1/(1-\widetilde{A})$ and
consider
\begin{equation*}\label{EqNf}
\omega_{\gamma\in(2,4),\theta\,}(t)=
\frac{1}{1-\widetilde{A}(1-itc_\theta)^\gamma}=\frac{1}{1+\widetilde{A}
\left[(1-itc_\theta)^{\widetilde{\gamma}}\right]^2}=\frac{a}{1-(1-a)
\left[(-izc_\theta)^{\widetilde{\gamma}}\right]^{2}}
\end{equation*}
with $z=t+i/c$. Clearly, the r.h.s. of the above can be represented
as $\mathcal{L}_L(-\log f_{\widetilde{\gamma}}^{stab}(z))$ where
$f_{\widetilde{\gamma}}^{stab}(z)=\exp\{A[-(-izc_\theta)^{\widetilde{\gamma}}]\}$
is the chf of one-sided stable random variable, as
$\mathcal{L}_L(u)=1/(1-b^2 u^2)$ is the Laplace transform of the
Laplace distribution. Clearly $\omega_{g}(z):=1/\left(1-(1-a)
\left[(-izc_\theta)^{\widetilde{\gamma}}\right]^{2}\right)$ is a chf
of geometric analogue of one-sided stable r.v. (provided that
$(a-1)>0$), and since $\omega_{g}(-i/c)=1/a$, then
$\omega_{\gamma\in(2,4),\theta\,}(t)=\omega_{g}(z-i/c)/\omega_{g}(-i/c)$.\,\,
Then $\omega_{\gamma\in(2,4),\theta\,}$ is itself a chf. Note that
the condition $(1-a)>0$ means that it should be that
$\widetilde{A}>1$.

When $\gamma>4$, we can argue analogously. Note that the r.v.
corresponding to the chf $\omega_{\gamma>2,\theta}$ is nevertheless
{\bf not} geometric stable, nor it is geometrically infinitely
divisible.
\end{itemize}

\section{Summary, interpretation and further extensions of the model}


\subsection{Summary of properties}

As the classification of different cases with respect to the ranges
of $\gamma$ spreads out to $f_{\gamma}$, $f_{\gamma,\theta}$ and
$\omega_{\gamma,\theta}$, it might be useful to summarize the
relevant information in a summary table.
\begin{table}[h!]
\begin{center}
\begin{tabular}{|p{1.4 cm}|p{3.0 cm}|p{3.4 cm}|p{3.8 cm}|p{2.2 cm}|}
\hline $\,$ & Case {\bf a} \par$\gamma<0$ & Case {\bf b}\par
$\gamma\in(0,1)$ & Case {\bf d}\par $\gamma\in(1,2)$ & Case {\bf
f}\par $\gamma>2$
\\\hline
$f_{\gamma}\,$ & $\,$ \par
$\exp\left\{\overline{A}\left[\widehat{\gamma}(t)-1\right]\right\}$
\par {\small $\bullet$ $A<0$}
& {\footnotesize with} $z:=t+i/c$\,:\par $\exp\left\{\,\,-[ -iz
c]^\gamma\,A\,\right\}$
\par {\small $\bullet$ $A>0$}
& {\footnotesize with} $z:=t+i/c$\,:\par $\exp\left\{-[ (-izc)^2
]^{\overline{\gamma}}\,A\right\}$
\par {\small $\bullet$ $A>0$}  & ---
\\\hline
$f_{\gamma,\theta}\,$ & $\,$ \par {\large
$e^{-\widetilde{A}\left[\widehat{\gamma}(t)-1\right]}$}
\par {\small $\bullet$ $\widetilde{A}<0$}  &
{\footnotesize with} $z:=t+i/c$\,:\par $\exp\left\{\,\,-[
-izc_\theta ]^\gamma\,\widetilde{A}\right\}$
\par {\small $\bullet$ $\widetilde{A}>0$}  &
{\footnotesize with} $z:=t+i/c_\theta$\,:\par $\exp\left\{-[
(-izc_\theta)^2 ]^{\overline{\gamma}}\,\widetilde{A}\right\}$
\par {\small $\bullet$ $\widetilde{A}>0$}
 & ---
\\\hline
\vspace{0.1cm} {\small Geometr.} $\omega_{\gamma,\theta}$ &
\vspace{0.01cm}
 $\,$\par
$\frac{1}{\alpha}\sum\limits_{n=0}^{\infty} \alpha^n f_{\gamma}^{n}$
\par {\small $\bullet$ $\widetilde{A}<0$}
& \vspace{0.01cm} {\small \par $(1-\ln
f_{\gamma\in(0,1),\theta}(t))^{-1}$}
\par {\small $\bullet$ $\widetilde{A}>0$}
& \vspace{0.01cm} \par $(1-\ln f_{\gamma\in(1,2),\theta}(t))^{-1}$
\par {\small $\bullet$ $\widetilde{A}>0$}
& \vspace{0.01cm} $ \frac{1-b}{1-b
\left[(-izc_\theta)^{\widetilde{\gamma}}\right]^2}$
\par {\small $\bullet$ $\widetilde{A}>1$}
\\\hline
\end{tabular}
\end{center}
\end{table}

\subsection{Interpretation in terms of a subordinated Levy process}

While in Section \ref{GSect} a new class was introduced whose chf
$\phi$ is linked with the chf $f_{\gamma,\theta}$ through
$\phi(t)=\mathcal{L}(- \ln f_{\gamma,\theta}(t))$, the chf
$f_{\gamma,\theta}$ itself, for $1<\gamma<2$, can be represented in
the above form. Indeed, according to Case {\bf d} of Paragraph
\ref{cases}, we have the following mixture representation for
$f_{\gamma\in(1,2)}$
\begin{equation}\label{LaplMix2}
f_{\gamma\in(1,2)}(t)= \int
\left[\phi_{(2,2\overline{c})}(t)\right]^xp_{\widetilde{\gamma}\in(0,1)}(x)dx=\int
e^{x\ln \phi_{(2,2\overline{c})}(t)}
p_{\widetilde{\gamma}\in(0,1)}(x)dx=\mathcal{L}_{\widetilde{\gamma}}(-
\ln \phi_{(2,2\overline{c})}(t)),
\end{equation}
and the same is valid for $f_{\gamma\in(1,2),\theta}$\,\,\,as well.

Recall that a r.v. $S$ with chf of the form $\mathcal{L}_N(- \ln
f_X(t))$ can be interpreted as a random sum (with random but finite
number of summands), if the Laplace transform $\mathcal{L}_N$
corresponds to a discrete random variable (a realization of a
counting process). One of the most celebrated examples is the
Poisson process whose Laplace transform is
$\exp\left\{\lambda\left(e^{-t}-1\right)\right\}$, so that the chf
$\exp\left\{\lambda\left(\phi(t)-1\right)\right\}$ corresponds to
the compound Poisson r.v. $S=\sum\limits_{i=1}^{N(t)}X_i$.

According to (\ref{1stmix}), in our case the LT of the r.v.
associated with the counting process is
$\mathcal{L}_{N}(u)=\exp\left\{A\left[1-(1+cu)^{\widetilde{\gamma}}\right]\right\}$
with $0<\widetilde{\gamma}<1$. Clearly, it corresponds to a {\it
continuous} r.v.; moreover, as discussed in Case {\bf b}, the chf
$f_{\widetilde{\gamma}\in(0,1)}$ of this r.v. is obtained from the
one-sided stable chf by a complex shift, s.t. the corresponding r.v.
relates to the exponential transformation of one-sided stable
density. Hence the chf (\ref{LaplMix2}) cannot be interpreted as a
chf of a compound r.v. in its usual form
$\sum\limits_{i=1}^{N(t)}X_i$.

However, the corresponding r.v. can be interpreted in terms of
continuous-time analogues of random sums --- the increments of a
{\it subordinated Levy process}. For the chf of any Levy process
$X_t$ on $\mathbb{R}$ we have (see e.g. \cite{ConT})
\[ \mathbf{E}\left[e^{izX_t}\right]=e^{t\psi_X(u)}\,,\,\,\,\,\,z\in\mathbb{R}\, \]
where $\psi_X$, usually called the {\it characteristic exponent} of
the Levy process, is defined via $\phi_X(u)=e^{\psi_X(u)}$\,, with
$\phi_X$ being the chf of the r.v. $X_1$\,.

Furthermore, for a subordinated Levy process $Y_t=X_{S(t)}$ with
subordinator $S(t)$ (which can be viewed is a random compression of
time) we have
\begin{equation}\label{chexp}
\mathbf{E}\left[e^{izY_t}\right]=e^{t\cdot\,
l_S\left(\psi_X(z)\right)}\,,\,\,\,\,\,z\in\mathbb{R}\,,
\end{equation}
where $l_S(z)$ defined via
$\mathbf{E}\left[e^{uS(t)}\right]=e^{t\cdot\,l_S(u)}$ is called the
{\it Laplace exponent} of $S$. For the chf of the increment $Y_1$ in
unit time, (\ref{chexp}) turns into $e^{l_S\left(\psi_X(z)\right)}$.

Recall that our chf $f_{\gamma\in(1,2)}$, according to
(\ref{LaplMix2}), can be represented as
\[
f_{\gamma\in(1,2)}(z)=\mathcal{L}_{\widetilde{\gamma}}(- \ln
\phi_{(2,2\overline{c})}(z))=
\exp\left\{A\left[1-(1-\ln\phi_{(2,2\overline{c})}(z))^{\widetilde{\gamma}}\right]\right\}=
e^{l_{{\widetilde{\gamma}}}\left(\ln\phi_{(2,2\overline{c})}(z)\right)},
\]
where $l_{\widetilde{\gamma}}$ is the Laplace exponent of the Levy
process whose increments have the chf
$f_{\widetilde{\gamma}\in(0,1)}$, i.e.
$f_{\widetilde{\gamma}\in(0,1)}(iu)=
\mathcal{L}_{\widetilde{\gamma}}(-u)=e^{l_{\widetilde{\gamma}}(u)}$.
Therefore $f_{\gamma\in(1,2)}$ can be interpreted as a chf of the
r.v. $Y_1$, i.e. of the increments of subordinated Levy process
$Y_t=X_{S(t)}$ in unit time $t=1$. The increments of the underlying
process $X_t$ have normal distribution with parameters
$(2,2\overline{c})$, while the increments of the subordinator $S(t)$
are distributed according to the exponentially transformed one-sided
Levy distribution whose chf is given by
$f_{\widetilde{\gamma}\in(0,1)}$.



\vspace*{0.5cm}

A possible way to extend the model is to use {\bf any} relevant chf
as the component $\phi$ in the representation (\ref{LaplMix2}). That
would also lead to an interpretation of the corresponding
distribution as a distribution of the increments of a subordinated
Levy process $Y_t=X_{S(t)}$, with the same subordinator $S(t)$ but
with a different underlying process $X_t$.

\appendix
$\,$\,\,\par
\section{Derivation of the chf in the explicit form}

The function $f$ to be found satisfies the balance equations
\begin{equation*}
f_\theta(t)=\frac{f(t-i\theta)}{f(-i\theta)}\,\,\,\,\mbox{and}\,\,\,\,f_\theta(t)=
f^{\alpha(\theta)}(\beta(\theta) t)\,\,\,\,\forall\,\,
t\in\mathbb{R},\,\,\,\,\, \theta\in[a,b]\,,\,\, 0\in[a,b].
\end{equation*}
In terms of the log-function $g(t)=\log f(t)$, it turns into
\begin{equation*}
g(t-i\theta)=g(-i\theta)+\alpha(\theta)g(\beta(\theta) t).
\end{equation*}

In order to prove that $g$ can be represented in the form
(\ref{gtSolut}), let us differentiate $g(t-i\theta)$ w.r.t. to $t$
and to $\theta$. First differentiate it w.r.t. to $t$
\begin{equation}\label{primtheta}
g^{\prime}(t-i\theta)=\alpha(\theta)\cdot\beta(\theta)\cdot
g^{\prime}(\beta(\theta)t).
\end{equation}
Considering a limit w.r.t $\theta\rightarrow0$ and assuming that
$\alpha(0)\beta(0)\neq0$, we get
\begin{equation}\label{diff4}
g^{\prime}(t)=Ag^{\prime}(bt),
\end{equation}
where $b:=\beta(0)$ and $A:=\alpha(0)\beta(0)$. Two major cases are
possible: $b\neq1$ and $b=0$. Let us start with the first one.

\paragraph{$b\neq1$}$\,$

Without loss of generality, assume that $\,|\,b\,|\,<1$. Let us
differentiate (\ref{diff4}) further $k$ times
\begin{equation*}
g^{(k+1)}(t)=A\cdot b^{k}g^{(k+1)}(bt),
\end{equation*}
and consider it at $t=0$
\begin{equation*}
g^{(k+1)}(0)=A\cdot b^{k}g^{(k+1)}(0).
\end{equation*}
Considering a limit w.r.t $k\rightarrow\infty$, we see that $1<A
b^k\rightarrow0$, which means that
\[ g^{(k+1)}(0)=0 \quad \mbox{when}\quad k>K_0\,\,\,\mbox{(\,for some large enough\,}\,K_0)\,. \]
It follows from the latter that $g(t)$ is some polynomial
$g(t)=P(t)$, so that the chf $f(t)$ is the exponent of a polynomial
$f(t)=e^{P(t)}$. Due to Marcinkiewicz's theorem \cite{Marcin}, we
can conclude that the degree of $P$ should not be higher than $2$,
e.g. it is a quadratic polynomial.

In other words, $b\neq1$ in (\ref{diff4}) corresponds to the case
when $f(t)$ is the chf of the Normal distribution considered in
Section 2.1 as Case e.

\paragraph{$b=1$}$\,$

Specifically, $b=1$ means that not just $\beta(0)=1$ but also
$\alpha(0)=1$, as follows from (\ref{primtheta}). Denoting
$\Lambda(\theta):=\alpha(\theta)\beta(\theta)$ and keeping in mind
that $\Lambda(0)=1$, let us now differentiate (\ref{primtheta})
w.r.t to $\theta$ which gives
\[
-ig^{\prime\prime}(t-i\theta)=\Lambda^{\prime}(\theta)g^{\prime}(\beta(\theta)t)+\Lambda(\theta)\beta^{\prime}(\theta)
g^{\prime\prime}(\beta(\theta)t)t
\]
and considering a limit w.r.t $\theta\rightarrow0$, we get
\[
-ig^{\prime\prime}(t)=\Lambda^{\prime}(0)g^{\prime}(t)+t\beta^{\prime}(0)g^{\prime\prime}(t),
\]
which leads to
\[
-(t\beta^{\prime}(0)+i)g^{\prime\prime}(t)=\Lambda^{\prime}(0)g^{\prime}(t)
\]
Denoting $u(t):=g^{\prime}(t)$, we get an equation

\begin{equation}\label{ut}
\frac{u^{\prime}(t)}{u(t)}=\frac{-\Lambda^{\prime}(0)}{(i+t\beta^{\prime}(0))}
\end{equation}
The solution of this equation gives
\[ \ln u(t)=-c_1\ln(i+t\beta^{\prime}(0))+c_2 \quad \Longleftrightarrow \quad
u(t)=\frac{c}{(t\beta^{\prime}(0)+i)^{c_1}}\,\,(\mbox{where}\,\,c:=e^{c_2})\,,
\] i.e. $g^{\prime}(t)=c(t\beta^{\prime}(0)+i)^{-c_1}$, which means
that
\[
g(t)=\frac{c}{(1-c_1)\beta^{\prime}(0)}\left(\beta^\prime(0)t+i\right)^{1-c_1}+c_2.
\]
Denote $1-c_1=:\gamma$ and, in order to express the constant $c_2$
explicitly, consider the above at $t=0$
\[
\frac{c}{(1-c_1)\beta^\prime(0)}i^\gamma=-c_2,
\]
so that we can denote $A_1:=\frac{c}{(1-c_1)\beta^\prime(0)}$ and
rearrange to get
\[ g(t)=A_1\left[(\beta^{\prime}(0)t+i)^\gamma-i^\gamma\right]. \]
With two more notations $A:=-A_1 i^\gamma$ and
$c:=\beta^{\prime}(0)$, after rearrangement it finally turns into
\[ g(t)=A\left[1-(1-ict)^\gamma\right]. \]

Throughout the derivation, we assumed that $\beta^{\prime}(0)\neq0$.
Let us now consider the opposite case.

\paragraph{$\beta^{\prime}(0)=0$} $\,$

It easy to check that this corresponds to a degenerate case, as
follows from (\ref{ut})
\[ \frac{u^\prime(t)}{u(t)}=-const=c\quad \Longleftrightarrow \quad \ln u(t) = ct + c_2, \]
which implies that $u(t)=\widetilde{c}e^{ct}.$

Eventually, the general form of the function $g$ in question is
\begin{equation*}
g(t)=A\left[1-(1-itc)^{\gamma}\right]\,.
\end{equation*}

\end{document}